\documentclass[10pt]{article}
\usepackage{setspace}
\usepackage{amsmath}
\usepackage{hyperref}
\usepackage{amssymb}
\usepackage{graphicx}
\usepackage{subfigure}
\usepackage{epstopdf}
\usepackage{achemso}

\def\la{\langle}
\def\ra{\rangle}

\def\blp{\left(}
\def\brp{\right)}

\def\fh{\mathcal{F}_h}
\def\fl{\mathcal{F}_l}

\begin{document}
\begin{center}
{\Large\bf Theory of biopolymer stretching at high forces}\\
\ \\
{\large Ngo Minh Toan$^1$ and D. Thirumalai$^{1,2}$}\\
\ \\
$^1$  Biophysics Program, Institute for Physical Science and Technology,\\
$^{1,2}$  Department of Chemistry and Biochemistry, University of Maryland at College Park, College Park, Maryland, USA, 20742
\end{center}

\doublespace
%\setstretch{2.0}
%\singlespace

\begin{abstract}
We provide a unified theory for the high force elasticity of biopolymers solely in terms of the persistence length, $\xi_p$, and the monomer spacing, $a$.  When the force $f>\fh \sim k_BT\xi_p/a^2$ the biopolymers behave as Freely Jointed Chains (FJCs) while in the range $\fl \sim k_BT/\xi_p < f < \fh$ the Worm-like Chain (WLC) is a better model. We show that $\xi_p$ can  be estimated from the force extension curve (FEC) at the extension $x\approx 1/2$ (normalized by the contour length of the biopolymer). After validating the theory using simulations, we provide a quantitative analysis of the FECs for a diverse set of biopolymers (dsDNA, ssRNA, ssDNA, polysaccharides, and unstructured PEVK domain of titin) for $x \ge 1/2$. The success of a specific polymer model (FJC or WLC) to describe the FEC of a given biopolymer is naturally explained by the theory. Only by probing the response of biopolymers over a wide range of forces can the $f$-dependent elasticity be fully described.
\end{abstract}

\section{Introduction}
The elasticity of biopolymers, probed using single molecule methods, is often analyzed using the worm-like chain (WLC)\cite{WLC1,WLC5,WLC3} model or the freely-jointed chain (FJC) model. For example, the force-extension curves (FECs) of double stranded (ds) DNA (at high monovalent salt concentration) are accurately reproduced using the WLC model with essentially the persistence length ($\xi_p$) as the only adjustable parameter while the FJC model does not fit the data\cite{busta}. The WLC model is also successful in describing the FEC of ssRNA~\cite{Koen} and proteins~\cite{linke,fernan}. In contrast, the FEC of ssDNA is better explained using the FJC model than the WLC model\cite{dessinges}. Given that the values of $\xi_p$ of ssDNA and ssRNA are similar~\cite{Caliskan05,Rivetti98}, it is hard to understand the source of the discrepancy. In addition, for polysaccharides~\cite{marsza1,marsza2,marsza3,marsza4} the FJC is used to extract the parameters from the FEC.

Here, we use general theoretical arguments to determine the polymer model that best describes the measured FEC. In particular, we derive a complete picture of the stretching physics of polymers at relative high forces, solely in terms of $\xi_p$ and the tensile screening length $\xi_t\equiv k_BT/f$~\cite{pincus}, where $k_BT$ is the thermal energy and $f$ is the stretching force. In the process, the FECs for the FJC and WLC models naturally appear as limiting cases within a unified theory for the stretching of polymer models, regardless of the nature of interactions between the monomers. Because the theory makes use of only general geometrical and scaling arguments that describe the response of a polymer under tension in terms of $\xi_p$ and $\xi_t$, we believe that our approach can be used to analyze the FECs of virtually any model polymer, and more importantly, many classes of biopolymers.

We propose that the FEC for any polymer should exhibit a near universal behavior. At very small forces we expect the extension to scale as $x\sim f$~\cite{flory,deGennes,ToanPhDThesis}. As the force increases the FEC could exhibit a plateau region (poor solvent)~\cite{Greg_Stretching07} or a universal behavior (good solvent) where the extension scales as $x\sim f^{2/3}$~\cite{pincus} (Pincus regime). Here we show that this behavior cannot extend beyond the region $x \sim 1/2$. In the force range $\fl \sim k_BT/\xi_p <f< \fh\sim k_BT\xi_p/a^2$ with $a$ being approximately the monomer size, the FEC should coincide with the prediction of the WLC model. For $f>\fh$ the polymer is better described using the FJC model. Thus, as long as $f>\fl$, the FEC can be fit using the analytical results from WLC or FJC model regardless of the interactions that stabilize the ordered structures of the biopolymer at low forces ($f<\fl$). The model that provides the most accurate fit of the measured FEC depends on the force range of the FEC as well as the specific biopolymer, which is characterized by $\xi_p$ and $a$. In other words, the elasticity of the biopolymer (or any polymer for that matter) depends on the force~\cite{dessinges}.

\section{Theory}
For a chain subject to a force $f$ along the $z$-axis, the tensile energy is $\mathcal{H}_S = -fa\sum_{i=1}^N\cos\theta_{f,i}$, where $\theta_{f,i}$ is the angle between bond $i$ and the force vector. In the limit the tensile screening length $\xi_t\equiv k_BT/f$\cite{pincus,deGennes} is less than $\xi_p$, the excluded volume or other solvent-mediated intrachain interactions are irrelevant\cite{ssDNAPincus,netz}. We make the physically reasonable assumption that, at high values of $f$, {\em the chain Hamiltonian has the global minimum when all the bond angles are zero}, which corresponds to the rod state. For chains with special bond constraints such as the freely rotating chain, we shall assume that we can define virtual bonds and virtual bond angles that satisfy this assumption.

\subsection{The FJC regime is universal for all discrete chains at high force}
In the following we denote $\la X\left(\theta\right) \ra_{\mathcal{H}_S} \equiv \int_0^\pi X(\theta) \exp\left(-{\mathcal{H}_S\over k_BT}\right)\sin\theta d\theta$ as the average with respect to $\mathcal{H}_S$ of $X$ as an arbitrary function of some angle $\theta$. For any two bonds $i$ and $j$ with bond angle $\theta_{ij}$, because of the triangle inequality $0 \le \theta_{ij} \le \theta_{f,i} + \theta_{f,j}$, the average of $\cos(\theta_{ij})$ with respect to {\em only} $\mathcal{H}_S$ satisfies:
\begin{eqnarray}\label{eq:costhetaij}
\la\cos\theta_{ij}\ra_{\mathcal{H}_S} &\ge& \la \cos\theta_{f,i}\cos\theta_{f,j}\ra_{\mathcal{H}_S} - \la \sin\theta_{f,i}\sin\theta_{f,j}\ra_{\mathcal{H}_S}\nonumber \\
&=&\la \cos\theta_{f,i}\ra_{\mathcal{H}_S}\la \cos\theta_{f,j}\ra_{\mathcal{H}_S} - \la \sin\theta_{f,i}\ra_{\mathcal{H}_S} \la \sin\theta_{f,i}\ra_{\mathcal{H}_S}\nonumber\\
&=&\blp \coth\left({a/\xi_t}\right) - {\xi_t\over a} \brp^2 - \blp {\pi\over 2} {I_1\blp{a/ \xi_t}\brp\over \sinh \blp{a/ \xi_t}\brp} \brp^2,
\end{eqnarray}
and becomes independent of $i$ and $j$. As the force becomes so large such that $\xi_t \ll a$, Eq.~\eqref{eq:costhetaij} for {\em any} two bonds reduces to
\begin{eqnarray}\label{eq:costhetaijf_expanded}
\la\cos\theta\ra_{\mathcal{H}_S} \gtrsim \exp\blp -{(\pi+4)\xi_t\over 2 a}\brp.
\end{eqnarray}
On the other hand, the contributions due only to the intra-chain interactions lead to
\begin{eqnarray}\label{eq:perlength_definition}
\la\cos\theta\ra_{\mathcal{H}_c}(s) &=&  \exp\blp-{s\over \xi_p}\brp,
\end{eqnarray}
where $\mathcal{H}_c$ is the chain Hamiltonian {\em of any form} and $s = |i-j|a$ is the arc-length. The expression in Eq.~\eqref{eq:perlength_definition}, which defines the persistence length, is assumed to be true for many types of uncharged biopolymers and for polyelectrolytes at sufficiently high counter-ion concentrations. At high forces, the contribution from the tensile energy to the average in Eq.~\eqref{eq:costhetaijf_expanded} becomes comparable to the contribution from the chain interaction. Thus, the arc-length $s$ at which the two contributions equal decreases as the force increases.
Eventually, as the force approaches $\fh$ such that $s=a$, we obtain
\begin{eqnarray}
\exp\blp -{(\pi+4)\xi_t\over 2 a}\brp \lesssim \la\cos\theta\ra_{\mathcal{H}_S}\blp \fh \brp = \exp\blp-{a\over \xi_p}\brp,
\end{eqnarray}
which leads to an estimate of $\fh$,
\begin{eqnarray}\label{eq:Fh}
\fh = c{k_BT \xi_p\over a^2},
\end{eqnarray}
with the constant $c$ in the range from ${4+\pi\over 2}$ to 4 (see below). For $s>a$, the average cosine is determined by the tensile energy $\mathcal{H}_S$ whose effects on the chain are (much) larger than intrachain interactions. At such high forces, the bonds align themselves along the force with small fluctuations. At forces above $\fh$, the stretching equation of a discrete semiflexible chain satisfying the above bond angle assumption becomes that of the FJC, i.e.
\begin{eqnarray}\label{eq:FJC}
x=1-{\xi_t\over a}.
\end{eqnarray}
Thus, $\fh$ should be viewed as a crossover force because above $\fh$ all biopolymers should behave as FJCs. Note that, in order for $\fh$ to be finite, $a$ in equation~\eqref{eq:Fh} must not vanish, i.e., the polymer chain should be viewed as a series of connected discrete links.

A similar form and notion of the crossover force were already pointed out elsewhere~\cite{angelo,Livadaru} for two specific models, WLC and freely-rotating chain (FRC). However, either the {\em explicit} form of the Hamiltonian (WLC) was required or simulations for FRC were needed to infer an empirical value of $\fh$. Here, general arguments based on the interplay between the force and intrinsic chain stiffness alone leads to an estimate of $\fh$ (Eq.\eqref{eq:Fh}). More importantly, $\fh$ can be obtained without assuming any specific energy function for the polymer, and hence should be considered as a universal behavior for any discrete semiflexible chain.

The critical value $\fh$ defined in eqn.~\eqref{eq:Fh} could, in principle, be somewhat more complicated especially if the bonds connecting the monomers are extensible. In this case, the stretching equation at high $f$ for the FJC (Eq.\eqref{eq:FJC}) has the form~\cite{WLC3,odijk1,nelson,Koen}:
\begin{eqnarray}\label{eq:FJC_K}
x=1-{\xi_t\over a} + {f\over K},
\end{eqnarray}
where $K$ is the stretch modulus that has the dimension of force.

\subsection{Force-dependent Kuhn segments and the universal effective FJC regime}
Consider the regime ${a^2\over c\xi_p} < \xi_t < \xi_p$ (the lower bound is when $f\approx \fh$). Similar to the Pincus argument at low force ($\xi_t > l_K$ - the Kuhn length) in which the chain can be viewed as a series of blobs, the polymer breaks up into a series of segments whose length is {\em on the order of} the Kuhn length and behaves {\em almost} similarly to a FJC of Kuhn segments (see Fig.~\ref{fig:regimes}). The relative extension in this force regime is, at most, $x \sim 1 - \xi_t/l_K$. However, we are not in the true FJC regime discussed above. The intrinsic intra-chain interactions, which tend to prevent the chain from bending, are relevant. The relative extension $x$ must be larger than that of the FJC with bond length $a$, namely $1-\xi_t/a$. Thus, we have the bound, $1 - {\xi_t\over a} \le x \le 1 - {\xi_t\over l_K}$. In analogy with Eq.~\eqref{eq:FJC}, we can formally write
\begin{eqnarray}
x = 1 - {\xi_t\over \lambda(f)},
\end{eqnarray}
where $\lambda(f)$ is the segment length of the effective FJC, which should be regarded as an effective force-dependent Kuhn length.

If the chain is  viewed as a sequence of connected segments with effective length $\lambda(f)$ that resembles the FJC, the average cosine of the angle formed by any two of these $\lambda$-segments is given by Eq.~\eqref{eq:costhetaijf_expanded} where $a$ is replaced by $\lambda$:
\begin{eqnarray}\label{eq:costheta_lambda_force}
\la \cos\theta\ra_{{\cal H}_S}(\lambda)& \gtrsim & \exp\blp -c{\xi_t \over \lambda}\brp.
\end{eqnarray}
The assumption that $\xi_t\ll \lambda$ has to be justified {\em a posteriori}. The intrinsic tangent-tangent correlation of any two $i$ and $j$ effective monomer-like $\lambda$-segments is
\begin{eqnarray}\label{eq:costheta_lambda_intrinsic}
\la \cos\theta\ra_{\mathcal{H}_b}(s) = \exp\blp -{|i-j|\lambda\over \xi_p}\brp.
\end{eqnarray}
By equating the two equations~\eqref{eq:costheta_lambda_force} and~\eqref{eq:costheta_lambda_intrinsic} (with $|i-j|=1$) we can relate $\lambda$, $f$ and $\xi_p$ using:
\begin{eqnarray}\label{eq:lambda_f_lp}
\lambda \approx \lambda_K \equiv \sqrt{c\xi_t\xi_p}.
\end{eqnarray}
It follows from Eq.\eqref{eq:lambda_f_lp} that the force-dependent Kuhn segment $\lambda_K$ is determined not only by the intrinsic stiffness of the chain through $\xi_p$ but also by the tension through $\xi_t$. The tensile screening length, which satisfies $\xi_t$($\propto f^{-1}$) $\ll$ $\lambda(f)$ ($\propto f^{-1/2}$) is similar to the {\em deflection length} proposed by Odijk~\cite{odijk1} using the explicit quadratic form of the Hamiltonian. The FEC in this regime satisfies,
\begin{eqnarray}\label{eq:2S}
x \approx 1 - \sqrt{{\xi_t\over c\xi_p}},
\end{eqnarray}
which coincides with the results for the WLC when $c=4$~\cite{WLC1,angelo,BYHa1997}. Thus, by merely using the analogy to a semiflexible chain, which behaves as a  FJC at high force and considering the effects of the intrinsic segment correlation and the effects of tension, we have obtained the characteristic stretching equation for {\em any} semiflexible chain. Such a regime exists in all semiflexible chains that satisfy the bond angle assumption. We shall refer to this as the effective FJC or WLC regime, generalizing the traditional name for any semiflexible chain. In practical applications the terms effective FJC and WLC can be used interchangeably.

\subsection{The unified stretching equation for both the FJC and WLC}
For the two regimes FJC and WLC discussed above, we assume that the generalized equation for $\lambda$ takes the following simple form,
\begin{eqnarray}\label{eq:lambda_general}
\lambda(f,\alpha) &\approx& \blp \lambda_K^{\alpha} + a^{\alpha} \brp^{1\over\alpha},
\end{eqnarray}
where $\alpha > 1$ is a model-dependent parameter. This equation reflects the observation that when $f <\fh$, $\lambda(f)$ is almost $\lambda_K$ ($\gg a$) and the chain behaves as an effective FJC; and when $f>\fh$, $\lambda(f)$ tends to $a$ ($\gg \lambda_K$) and the chain behaves as a true FJC. Thus, we have the equation for the two strong stretching regimes:
\begin{eqnarray}\label{eq:general_1S_2S}
x \approx 1 - {\xi_t\over \blp{\lambda_K^\alpha + a^\alpha}\brp^{1\over\alpha}},
\end{eqnarray}
which reduces exactly to the form for the discrete WLC~\cite{angelo} when $\alpha = 2$.

Because of the upper limit of $\lambda$, which is $l_K=2\xi_p$, the lower limit on the force for observing the effective FJC regime (assuming $c\lesssim 4$) is:
\begin{eqnarray}\label{eq:Fl}
\lambda = l_K \Rightarrow \xi_t \approx \xi_p \Rightarrow \fl \approx {k_BT\over \xi_p}.
\end{eqnarray}
Here, $\fl$ can be called the crossover force to the effective FJC regime from a low force regime (see Fig.~\ref{fig:regimes}).

The physics of the stretching of all biopolymer molecules using discrete semiflexible chains is summarized in Fig.\ref{fig:regimes}. At very low forces, the extension increases linearly with the force~\cite{ToanPhDThesis}. For larger forces, but not too large, such that the tensile screening length is still significantly larger than the Kuhn length, one reaches the Pincus regime of blobs where the extension scales either linearly with the force (ideal chains) or as $f^{2/3}$ (self-avoiding chains)\cite{pincus,deGennes}. As the force increases until it is larger than $\fl$, $\xi_t$ becomes smaller than the Kuhn length, the chain behaves as an effective FJC with the law $1-x \propto f^{-1/2}$. Eventually, when $\xi_t$ becomes smaller than $a^2/\xi_p$ ($f>\fh$), the chain behaves as a true FJC under high tension with the FEC law $1-x \propto f^{-1}$.

\section{Simulations confirm the theoretical predictions}\label{sec:simulations}
To test our theory, we performed simulations using a number of models.

{\em ``Polynomial'' models (``Generalized'' WLC).} The Hamiltonian of the chain with $N$ bonds is
\begin{eqnarray}\label{eq:GeneralizedWLC_Hamiltonian}
{\cal H}_b &=& {\kappa_b\over 2}\sum_{i=1}^N \blp{\theta\over \Theta}\brp^\gamma,
\end{eqnarray}
where $\theta$, the angle between two successive bonds, varies from $0$ to $\pi$. When $\gamma=2$ and $\Theta = 1$, the Hamiltonian approximately reduces to the WLC energy function and  when $\gamma$ is large, we recover the local approximation Thick Chain model~\cite{BiophysJ2005,ToanJPCM06}. Fig.~\ref{fig:sub_PP} shows the dependence of $\mathcal{H}_b$ on $\theta/\Theta$ for several values of $\gamma$.

{\em The FBNE (finite bendable non-linear elastic) chain.} To model the finite bendability of a chain, we use the FBNE potential
\begin{eqnarray}\label{eq:FBNE_Hamiltonian}
\mathcal{H}_{FBNE} &=&\left\{
                   \begin{array}{ll}
                     -{\kappa_b\over 2}\sum_{i=1}^N\log\blp 1 - \blp{\theta\over\Theta}\brp^2\brp, & \hbox{$\theta<\Theta$;} \\
                     \infty, & \hbox{$\theta \ge\Theta$,}
                   \end{array}
                 \right.
\end{eqnarray}
which resembles the FENE potential~\cite{kremer1,SOP} for chain bonds.

To compare all simulation data, we chose parameters in the models such that the persistence length is fixed at $\xi_p/a = 10$. The ratio
${\fh/\fl} \approx c\left({\xi_p/ a}\right)^2 = 100 c$ is large enough to clearly observe the  effective FJC regime. From equations~\eqref{eq:2S} and~\eqref{eq:FJC}, it can be seen that plotting $1-x$ against $f\times a/k_BT$ in double logarithmic scale would give us the characteristic slopes of $-1/2$ and $-1$ for the effective FJC and FJC regimes, respectively. Indeed, Figs.~\ref{fig:sub_log1xlog1fsimulations1} and~\ref{fig:sub_log1xlog1ffit} shows that there are these characteristic slopes for the ``polynomial'' model at all $\gamma$ considered. The same result is obtained for the FBNE and {\em self-avoiding} TC models~\cite{BiophysJ2005,ToanJPCM06,ToanPhDThesis} (data not shown). Moreover, the fits using the generalized equation~\eqref{eq:general_1S_2S} with {\em two} free parameters $c$ and $\alpha$ are excellent and yield, in accord with theoretical arguments, $c$ in the range ${4+\pi\over 2}$ to $4$, and $\alpha$ in the range $2$ to $7$ and positively correlated with $\gamma$. Moreover, it is easily seen that the effective FJC regime sets in as soon as $f \approx 0.1 {k_BT\over a} = {a\over \xi_p}{k_BT\over a} \equiv \fl$. The crossover from effective FJC to FJC occurs at  $ f\approx 40 {k_BT\over a} \approx {c\xi_p\over a} {k_BT\over a} \equiv \fh$. Thus, the simulations confirm the theoretical arguments.

\section{Analysis of experiments}

\subsection{Significance of the regimes and crossover forces}
Our analysis provides a theoretical basis for choosing the appropriate polymer model to analyze the FECs of biopolymers. In order to choose a specific model it is necessary to estimate $\fl$ and $\fh$, which require the values of $a$ and $\xi_p$ (see eqs.~\eqref{eq:Fl} and~\eqref{eq:Fh}) that are intrinsic properties of the biopolymer. The value of the monomer spacing $a$ can be obtained by analyzing the FEC at high force (FJC regime). The effects of the self-avoidance of the chain can be assessed most accurately in the linear and Pincus regimes (Fig.\ref{fig:regimes}). Chain extensibility can make it difficult to estimate $a$ making it necessary to include models that treat chain extensibility explicitly~\cite{WLC3,odijk1,nelson}. The persistence length is easily extracted from the FECs in the effective FJC regime. It is the most accurate quantity that can be inferred from experiments, since it is the effective FJC regime that is commonly detectable in those experiments~\cite{busta,WLC5,strick,WLC3,Koen,linke,linke2,fernan,fernan2}. In practice, we use the force at $x\approx 0.5$ (referred to as 1/2-rule) to estimate $\fl = k_BT/\xi_p$, and hence $\xi_p$ (see Eq.~\eqref{eq:2S} with $c\approx 4$). We illustrate the utility of our theory that depends on the interplay of the two ``crossover'' forces $\fl$ and $\fh$ in determining the elasticity of biopolymers by analyzing measured FECs.

{\bf dsDNA}: Using the well-known values $\xi_p \approx 50$nm~\cite{WLC1,WLC4,WLC5,WLC3} and $a = 0.34$ nm at $T = 300$K, we obtain $\fl \approx 0.1$pN and $\fh \approx 6400 $pN. Values of $f$ that are comparable to $\fl$ are accessible only using magnetic beads or optical tweezers~\cite{busta,WLC5,strick,WLC3}, whereas $f \sim \fh$ can be realized in AFM experiments. Because dsDNA undergoes a phase transition to an overstretching form at $f\approx 65$ pN~\cite{plengthssdna,single2,Bustamante2000,bustamante03}, the FJC limit cannot be observed in dsDNA. It should be stressed that even if $a$ is greatly increased, up to the conceivably maximum value of 3.4 nm or one double helix turn, $\fh$ still exceeds $f\approx 65$pN. These estimates show that the FEC of dsDNA (excluding the overstretching region) should be in only the effective FJC regime, which explains the success of the WLC in quantitatively describing the FEC of dsDNA.

{\bf ssRNA}: From the typical values of $\xi_p\approx 1$nm\cite{Caliskan05,Rivetti98} at sufficiently high counter-ion concentrations and $a \approx 0.55$ nm~\cite{Hyeon2006} for both ssRNA and ssDNA we obtain $\fl \approx 4$ pN and $\fh\approx 50$ pN. Thus, both $\fl$ and $\fh$ are small enough that they can be accessed in typical LOT experiments, making it practical to infer the $f$-dependent changes in the elasticity. The values of $\fl$ and $\fh$ readily explain why the measured FEC for poly-U using optical tweezers~\cite{Koen} in the force range from about 0.5 pN to 50 pN, which is about $\fh$, is best described using the WLC model. We predict that the FJC behavior should emerge for forces larger than 50 pN for poly-U. The same arguments also hold good for ssDNA data at high ionic concentrations in the following paper by Saleh et al.

{\bf ssDNA}: In a recent experiment, Saleh et al. measured FEC of ssDNA at varying counter-ion concentrations in the force range $0.1\lesssim f \lesssim 50$pN~\cite{ssDNAPincus}. The data suggests that for $f<f_c$, a critical force that depends on the salt concentration, the FEC is in the Pincus regime (Fig.~\ref{fig:regimes}) in which the non-linear response to force is determined largely by the excluded-volume interactions. We adopt the $1/2$-rule to calculate $\xi_p(\lambda_D)$ as a function of the Debye length $\lambda_D$ for ssDNA and poly-U. The estimated $\xi_p(\lambda_D)$ for these two polyelectrolytes are identical (Fig.~\ref{fig:fit_ssRNA_DNA}). Moreover, we can fit the curves with  $\xi_p(\lambda_D) = \xi_p^0 + A\lambda_D$, where $\xi_p^0$ is the ``bare'' persistence length and $A$ is a proportionality constant. The values of $A$ and $\xi_p^0$ for poly-U and ssDNA are nearly identical ($\xi_p^{0,\text{poly-U}} = 0.67$nm and $\xi_p^{0,\text{ssDNA}} = 0.63$nm). The linear dependence of $\xi_p$ on $\lambda_D$ is in agreement with previous works~\cite{Barra93,Dobrynin05} as well as Saleh's findings~\cite{ssDNAPincus}. Although the procedure used by Saleh et al. to extract $\xi_p(\lambda_D)$ is different from the WLC-based analysis presented here, our estimated $\xi_p^{0,\text{ssDNA}}$ is in excellent agreement with their estimate of 0.6nm~\cite{ssDNAPincus}.

Comparison of our predictions with the FEC data for ssDNA reported by Rief et al.~\cite{MRief} in the force range from 10pN to 200pN and in Tris buffer further validates the present analysis. In a different paper~\cite{dessinges}, Dessinges et al. combined the data with their data on ssDNA in the force range from 0.05pN to 20pN and in a different buffer (see Fig. 4 in ref.~\cite{dessinges}). Using extensive data on ssDNA, they showed that at very low force, self-avoidance and electrostatic effects play dominant roles, and hence the elasticity of ssDNA is qualitatively different from the high force regimes. Indeed, by taking into account the electrostatic interactions in the so-called modified (extensible) FJC model, Dessinges et al. explained the low force part of the FEC. They also showed that {\em both} the WLC and the ``bare'' modified FJC, when applied independently, failed to fit the {\em entire} high force part ($x \ge 1/2$) of the FEC. Here we show that the FEC above $f=\fl$ (data by Rief et al.~\cite{MRief}) can be quantitatively analyzed using our theory. The equation for the high stretch regimes (Eq.~\eqref{eq:general_1S_2S}), which has both the WLC and FJC high force behaviors, provides a very good fit to the data. We notice that the modified FJC model by Dessinges et al. with electrostatic interactions actually failed to fit the high force ($x>1/2$) portion of the FEC~\cite{dessinges}. It could be argued that the good fit to the lower portion of the FEC was mainly because the electrostatic interactions, which play dominant roles, were well taken into account. The fact that whether the FJC or WLC is used for the low force regime should not considerably affect the outcome of the fit. But for the high force regime, it is essential to do a force range analysis prior to the actual fit, as it is done here. Our extracted values are $a\approx 0.5$ nm, $\xi_p\approx 0.8$ nm and $\alpha \approx 7$ leading to $\fl \approx 5$ pN and $\fh \approx 55$ pN (see Fig.\ref{fig:fit_dessinges}).

{\bf Protein (PEVK domain)}: PEVK (Pro-Glu-Val-Lys) domain is a proline-rich domain of the muscle protein titin and it has no definite folded structure. Using the accepted values $\xi_p \approx 0.5$nm~\cite{linke,linke2,fernan,fernan2} and $a = 0.38$ nm, $\fl \approx 8.4$ pN, which is well in the covered range of almost all single molecule apparatuses, and $\fh \approx 50$ pN. We expect that the WLC is more suitable for the PEVK domain than the FJC, and indeed FECs in the force range from 10 to 200pN were analyzed using WLC~\cite{linke,linke2,fernan,fernan2}. However, we can see the need for a discrete model rather than the continuum WLC to obtain a better fit.

{\bf Polysaccharides}: Linear polysaccharides that occur naturally in a variety of cellular structures are subject to tensile stress. At high forces the sugar rings undergo a transition from chair to boat conformation, which clearly changes their elastic properties~\cite{marsza2,marsza3}. Unlike other systems (dsDNA, for example) the change in elasticity is associated with enthalpic changes as a result of the conformational transition. Surprisingly, it is found that over a wide range of forces the FEC is best explained by the FJC model~\cite{marsza4}, which can be fully explained using our theory. For cellulose with the value $l_K\approx a\approx 0.54$nm~\cite{marsza1,marsza2,marsza3,marsza4}, we obtain $\fl \approx \fh \approx 15.3$pN. This means that excluding the overstretch range, the entire range of FEC for cellulose reported in references by Marszalek et al.~\cite{marsza1,marsza2,marsza3,marsza4} obtained using AFM with force range of 10 to 1000 pN is best described using the FJC. The same observation can be made for other polysaccharides, such as dextran ($l_K \approx a \approx 0.44$ nm in the chair conformation and $l_K \approx a \approx 0.57$ nm in the boat conformation) and amylose ($l_K \approx a \approx 0.45$ nm (chair) and $l_K \approx a \approx 0.54$ nm (boat))~\cite{marsza2}. Thus, the intrinsic properties of polysaccharides show that this class of molecules behave as FJCs.

In all the above examples, it is necessary to assess the range of force in any particular FEC before using the polymer model that can best describe the data. In most cases, it is necessary to consider both the semi-flexibility and discreteness of the biopolymer. Models such as the discrete WLC model\cite{angelo} and the TC model~\cite{BiophysJ2005,ToanJPCM06} are better for most fitting purposes.
In analyzing most of the currently available data we find that these two models are adequate. However, as shown by Dessinges et al.~\cite{dessinges} the variations in the elasticity of ssDNA over a wide range of forces will require a combination of models. Our theory accounts for $f$-dependent variations in the elasticity in terms of naturally occurring length scales as long as $x\ge 1/2$. Below $x=1/2$, additional forces may stabilize specific structures.

\subsection{Overstretching in a FEC}
For each biopolymer, there is a threshold force at which the transition to bond overstretching occurs. Inextensible models can be used to fit only the experimental FECs below the threshold force. From the approximate equation \eqref{eq:general_1S_2S} in the high stretch regime it can be shown that the slope of the FEC above $\fl$ in the log-linear scale is a strictly monotonically increasing quantity for a generic semiflexible chain. On the other hand, in the region where overstretching effect is significant, the extension can be approximated as~\cite{odijk1} $x \approx 1 - {C\over f^{1/m}} + {f\over K}$, where $K$, $m$ and $C$ are positive constants. It can be shown that the rate of change of the slope in the region $x>1$ (overstretched) is always negative. Then for a FEC with overstretching, the threshold force is the one ($>\fl$) at which the slope of the curve $\ln f(x)$ vs. $x$ stops increasing. A slope that stays constant or decreases would signify the failure of the inextensible model.

The simple analysis, that accounts for extensibility through the stretch modulus $K$, allows us to gauge the threshold force from the FEC. If the data are precise, we can take the numerical derivative of $\ln f(x)$ in the region $f>\fl$ and detect the force at which it reaches a maximum value or a plateau. Otherwise, the practical solution would be to plot $f(x)$ vs $x$ in the log-linear plane then detect the change in the slope above $\fl$ by inspection. Although not rigorous, it turns out to be an effective way as demonstrated in Figure~\ref{fig:threshold_detection}. For the poly-U data~\cite{Koen} at various salt concentrations and at forces $f \le 50$pN, , the practical technique reveals no decrease in the slope in the $x>1/2$ region (data not shown). Therefore, we expect no overstretching in poly-U or ssDNA.

\section{Conclusions}
Our theory shows that the WLC and FJC models at large forces are the two limiting universal behaviors for virtually all polymers and especially biopolymers. The crossover from WLC to FJC is determined by the persistence length $\xi_p$ and the monomer spacing $a$. The validity of the theory is strongly demonstrated through the excellent agreement with simulation results and the successful analysis of the force extension curves of a diverse set of biopolymers. The two crossover forces, $\fh$ and $\fl$, which demarcate the boundary between the applicability of the WLC and FJC, provide a physical picture of the $f$-dependent elasticity of biopolymers. The derived $1/2$-rule is a convenient way to obtain a good estimate of the persistence length of the biopolymer from the measured FEC, which in turn yields the model that is most appropriate for analyzing the FEC. The theory also shows that complete understanding of the elasticity of biopolymers requires measurement of FECs for a wide range of forces as obtained for ssDNA~\cite{dessinges}.

{\bf Acknowledgements:} We are grateful to Changbong Hyeon, Greg Morrison and Omar Saleh for useful discussions. This work was supported by a grant from the National Science Foundation (CHE 09-14033).

\singlespace
%\bibliographystyle{biophysj}
%\bibliography{highstretch}

\begin{thebibliography}{10}

\bibitem{WLC1}
Marko,~J.~F.; Siggia,~E.~D. \emph{Macromolecules} \textbf{1995}, \emph{28},
  8759--8770.

\bibitem{WLC5}
Wang,~M.~D.; Yin,~H.; Landick,~R.; Gelles,~J.; Block,~S.~M. \emph{Biophys. J.}
  \textbf{1997}, \emph{72}, 1335--1346.

\bibitem{WLC3}
Bouchiat,~C.; Wang,~M.~D.; Allemand,~J.~F.; Strick,~T.; Block,~S.~M.;
  Croquette,~V. \emph{Biophys. J.} \textbf{1999}, \emph{76}, 409--413.

\bibitem{busta}
Smith,~S.~B.; Finzi,~L.; Bustamante,~C. \emph{Science} \textbf{1992},
  \emph{258}, 1122--1126.

\bibitem{Koen}
Seol,~Y.; Skinner,~G.; Visscher,~K. \emph{Phys. Rev. Lett.} \textbf{2004},
  \emph{93}, 118102.

\bibitem{linke}
Linke,~W.~A.; Kulke,~M.; Li,~H.; F.-Becker,~S.; Neagoe,~C.; Manstein,~D.~J.;
  Gautel,~M.; Fernandez,~J.~M. \emph{J. Struct. Biol.} \textbf{2002},
  \emph{137}, 194--205.

\bibitem{fernan}
Li,~H.; Oberhauser,~A.~F.; Redick,~S.~D.; C.-Vazquez,~M.; Erickson,~H.~P.;
  Fernandez,~J.~M. \emph{Proc. Natl. Acad. Sci. USA} \textbf{2001}, \emph{98},
  10682--10686.

\bibitem{dessinges}
Dessinges,~M.-N.; Maier,~B.; Zhang,~Y.; Peliti,~M.; Bensimon,~D.; Croquette,~V.
  \emph{Phys. Rev. Lett.} \textbf{2002}, \emph{89}, 248102.

\bibitem{Caliskan05}
Caliskan,~G.; Hyeon,~C.; Perez-Salas,~U.; Briber,~R.~M.; Woodson,~S.~A.;
  Thirumalai,~D. \emph{PRL} \textbf{2005}, \emph{95}, 268303.

\bibitem{Rivetti98}
Rivetti,~C.; Walker,~C.; Bustamante,~C. \emph{JMB} \textbf{1998}, \emph{280},
  41--59.

\bibitem{marsza1}
Marszalek,~P.~E.; Pang,~Y.-P.; Li,~H.; Yazal,~J.~E.; Oberhauser,~A.~F.;
  Fernandez,~J.~M. \emph{Proc. Natl. Acad. Sci. USA} \textbf{1999}, \emph{96},
  7894--7898.

\bibitem{marsza2}
Marszalek,~P.~E.; Oberhauser,~A.~F.; Pang,~Y.-P.; Fernandez,~J.~M.
  \emph{Nature} \textbf{1998}, \emph{396}, 661--664.

\bibitem{marsza3}
Marszalek,~P.~E.; Li,~H.; Fernandez,~J.~M. \emph{Nature Biotech.}
  \textbf{2001}, \emph{19}, 258--262.

\bibitem{marsza4}
Lee,~G.; Nowak,~W.; Jaroniec,~J.; Zhang,~Q.; Marszalek,~P. \emph{Biophys. J.}
  \textbf{2004}, \emph{87}, 1456--1465.

\bibitem{pincus}
Pincus,~P. \emph{Macromolecules} \textbf{1976}, \emph{9}, 386--388.

\bibitem{flory}
Flory,~P. \emph{Statistical mechanics of chain molecules};
\newblock Hanser Publisher: Munich, 1989.

\bibitem{deGennes}
de~Gennes,~P.-G. \emph{Scaling Concepts in Polymer Physics};
\newblock Cornell University Press: Ithaca, 1979.

\bibitem{ToanPhDThesis}
Toan,~N.~M. \emph{{PhD} Thesis}, SISSA: Trieste, Italy, 2006.

\bibitem{Greg_Stretching07}
Morrison,~G.; Hyeon,~C.; Toan,~N.~M.; Ha,~B.~Y.; Thirumalai,~D.
  \emph{Macromolecules} \textbf{2007}, \emph{70}, 7343--7353.

\bibitem{ssDNAPincus}
Saleh,~O.~A.; McIntosh,~D.; Pincus,~P.; Ribeck,~N. \emph{Phys. Rev. Lett.}
  \textbf{2009}, \emph{102}, 068301.

\bibitem{netz}
Netz,~R.~R. \emph{Macromolecules} \textbf{2001}, \emph{34}, 7522--7529.

\bibitem{angelo}
Rosa,~A.; Hoang,~T.~X.; Marenduzzo,~D.; Maritan,~A. \emph{Macromolecules}
  \textbf{2003}, \emph{36}, 10095--10102.

\bibitem{Livadaru}
Livadaru,~L.; Netz,~R.~R.; Kreuzer,~H.~J. \emph{Macromolecules} \textbf{2003},
  \emph{36}, 3732--3744.

\bibitem{odijk1}
Odijk,~T. \emph{Macromolecules} \textbf{1995}, \emph{28}, 7016--7018.

\bibitem{nelson}
Storm,~C.; Nelson,~P.~C. \emph{Phys. Rev. E} \textbf{2003}, \emph{67}, 051906.

\bibitem{BYHa1997}
Ha,~B.~Y.; Thirumalai,~D. \emph{J. Chem. Phys.} \textbf{1997}, \emph{106},
  4243--4247.

\bibitem{BiophysJ2005}
Toan,~N.~M.; Marenduzzo,~D.; Micheletti,~C. \emph{Biophys. J.} \textbf{2005},
  \emph{89}, 80--86.

\bibitem{ToanJPCM06}
Toan,~N.~M.; Micheletti,~C. \emph{J. Phys.: Condens. Matter} \textbf{2006},
  \emph{18}, S269--S281.

\bibitem{kremer1}
Kremer,~K.; Grest,~G.~S. \emph{J. Chem. Phys.} \textbf{1990}, \emph{92},
  5057--5086.

\bibitem{SOP}
Hyeon,~C.; Dima,~R.~I.; Thirumalai,~D. \emph{Structure} \textbf{2006},
  \emph{14}, 1633--1645.

\bibitem{strick}
Strick,~T.~R.; Allemand,~J.-F.; Bensimon,~D.; Bensimon,~A.; Croquette,~V.
  \emph{Science} \textbf{1996}, \emph{271}, 1835--1837.

\bibitem{linke2}
Li,~H.; Linke,~W.; Oberhauser,~A.; Carrion-Vazquez,~M.; Kerkvliet,~J.; Lu,~H.;
  Marszalek,~P.; Fernandez,~J. \emph{Nature} \textbf{2002}, \emph{418},
  998--1002.

\bibitem{fernan2}
Fisher,~T.~E.; Oberhauser,~A.~F.; Carrion-Vazquez,~M.; Marszalek,~P.~E.;
  Fernandez,~J.~M. \emph{TIBS} \textbf{1999}, \emph{24}, 379--384.

\bibitem{WLC4}
Baumann,~C.~G.; Smith,~S.~B.; Bloomfield,~V.~A.; Bustamante,~C. \emph{Proc.
  Natl. Acad. Sci. USA} \textbf{1997}, \emph{94}, 6185--6190.

\bibitem{plengthssdna}
S.~B.~Smith,~Y.~J.~C.; Bustamante,~C. \emph{Science} \textbf{1996}, \emph{271},
  795--799.

\bibitem{single2}
Cluzel,~P.; Lebrun,~A.; Heller,~C.; Lavery,~R.; Viovy,~J.-L.; Chatenay,~D.;
  Caron,~F. \emph{Science} \textbf{1996}, \emph{271}, 792.

\bibitem{Bustamante2000}
Wuite,~G.~J.; Smith,~S.~B.; Young,~M.; Keller,~D.; Bustamante,~C. \emph{Nature}
  \textbf{2000}, \emph{404}, 103--106.

\bibitem{bustamante03}
Bustamante,~C.; Bryant,~Z.; Smith,~S.~B. \emph{Nature} \textbf{2003},
  \emph{421}, 423--427.

\bibitem{Hyeon2006}
Hyeon,~C.; Dima,~R.~I.; Thirumalai,~D. \emph{J. Chem. Phys.} \textbf{2006},
  \emph{125}, 194905.

\bibitem{Barra93}
Barrat,~J.-L.; Joanny,~J.-F. \emph{Europhys. Lett.} \textbf{1993}, \emph{24},
  333--338.

\bibitem{Dobrynin05}
Dobrynin,~A.~V. \emph{Macromolecules} \textbf{2005}, \emph{38}, 9304--9314.

\bibitem{MRief}
Rief,~M.; Clausen-Schaumann,~H.; Gaub,~H.~E. \emph{Nature} \textbf{1999},
  \emph{6}, 346--349.

\end{thebibliography}
\providecommand{\url}[1]{\texttt{#1}}
\providecommand{\refin}[1]{\\ \textbf{Referenced in:} #1}

%\setstretch{2.0}
\newpage
\section*{Figure captions}

\begin{figure}[h]
  \centering
  \includegraphics[width=6in]{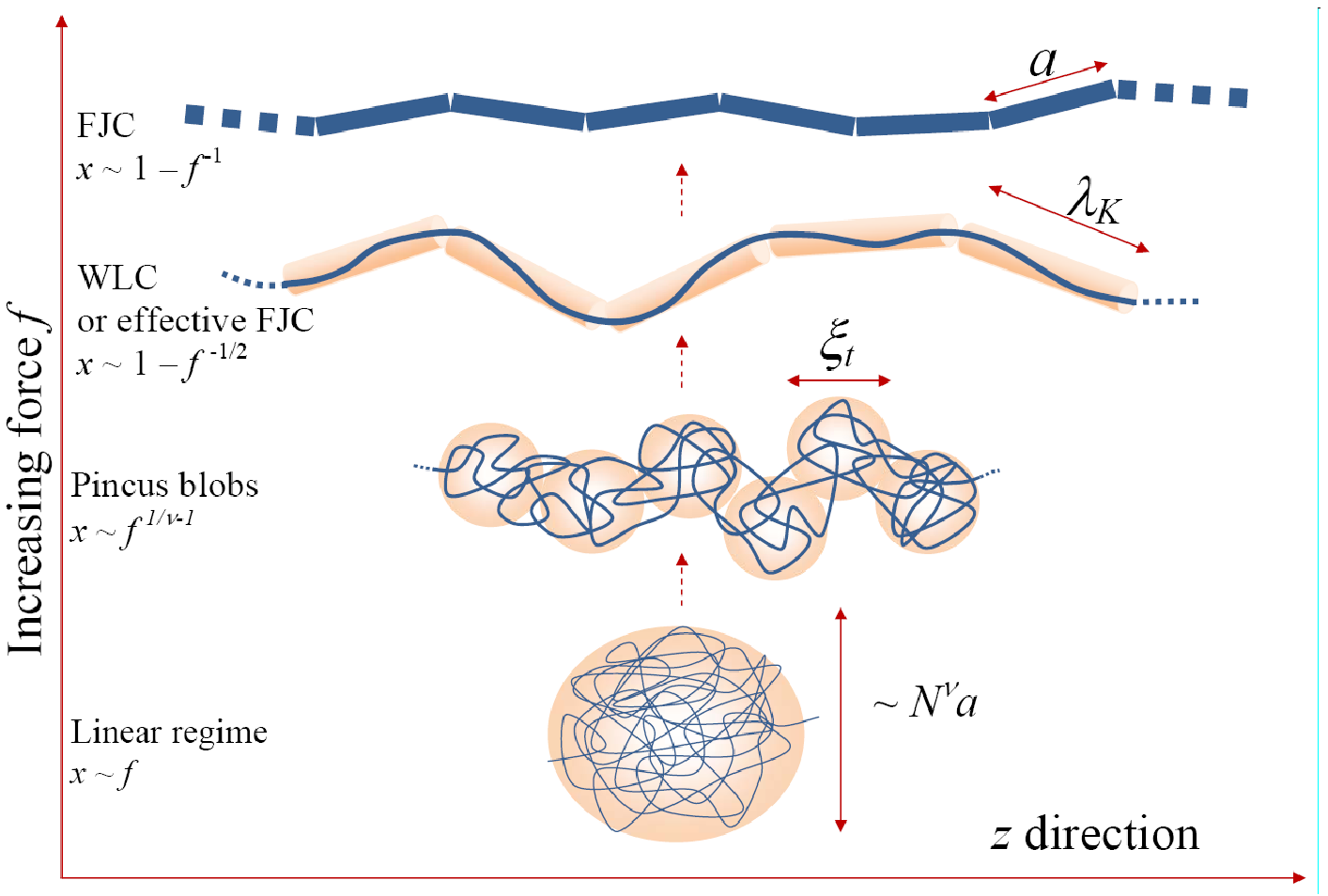}
  \caption{}
  \label{fig:regimes}
\end{figure}

{\bf Figure \ref{fig:regimes}}. Summary of the various stretching regimes in polymers or biopolymers. In the smallest force regime, the extension increases linearly with the force. In the case of biopolymers (RNA and protein) the structure in the collapsed state is ordered. Polymers in poor solvents are collapsed. The size of the polymer or a globular protein is about $N^\nu a$, where $N$ is the number of monomers, $a$ is the monomer spacing and $\nu$ is the scaling exponent, which is $\approx 1/3$ in poor solvent, $1/2$ in $\theta$-solvent or for ideal chains, and $3/5$ in good solvent~\cite{deGennes}. In good or $\theta$-solvent, as the force increases the chain extends into a series of Pincus blobs with typical size $\xi_t = k_BT/f$. In this regime, the extension scales as $x\propto f^{1/\nu-1}$~\cite{pincus,deGennes}. When the force is comparable to $\fl = k_BT/\xi_p$ ($\xi_t \approx \xi_p$), each Pincus ``blob'' contains just a single segment on the order of a persistence length. In this force range, the Pincus scaling law breaks down and the elasticity if described by the WLC model. The chain can be viewed as an effective freely jointed chain of segments of length $\lambda_K\sim \sqrt{\xi_t\xi_p}$. Intrachain interactions may still be relevant within the length scale $\lambda_K$ but are negligible on larger length scales. The extension goes as $1-x \propto f^{-1/2}$. Finally, when the force reaches $\fh \sim k_BT\xi_p/a^2$, all chain bonds are strongly aligned with the force and the intrinsic chain interactions are irrelevant. Thus, the chain behaves as a freely jointed chain (segment length $a$) with the extension scaling as $1-x\propto f^{-1}$. In some cases of biopolymers discussed in the main text, $\fh$ can be so high such that bond overstretching can occur before the force reaches $\fh$ and the theoretical FJC regime may not be actually observed.

\begin{figure}[h]
  \centering
  \subfigure[]
  {
    \includegraphics[width=2.6in]{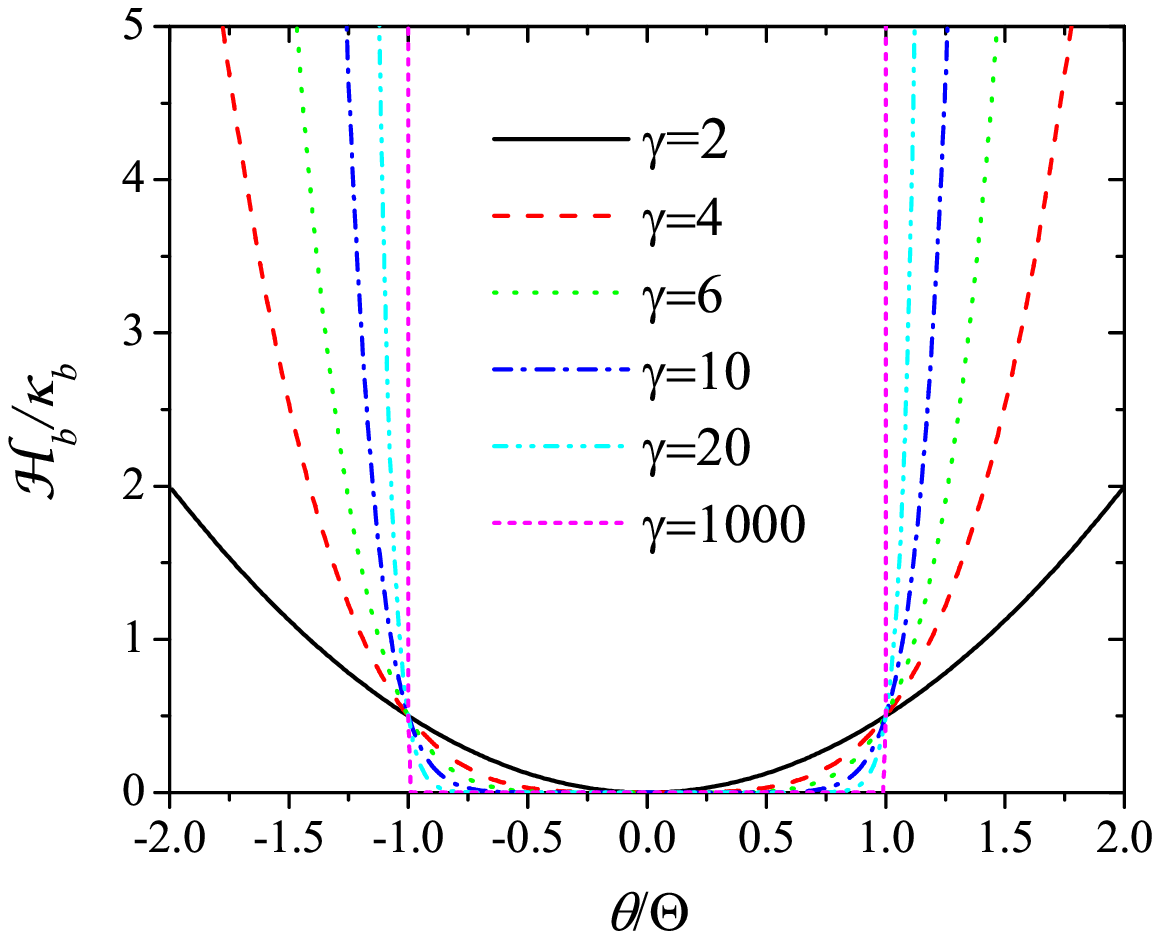}
    \label{fig:sub_PP}
  }
  \subfigure[]
  {
    \includegraphics[width=2.6in]{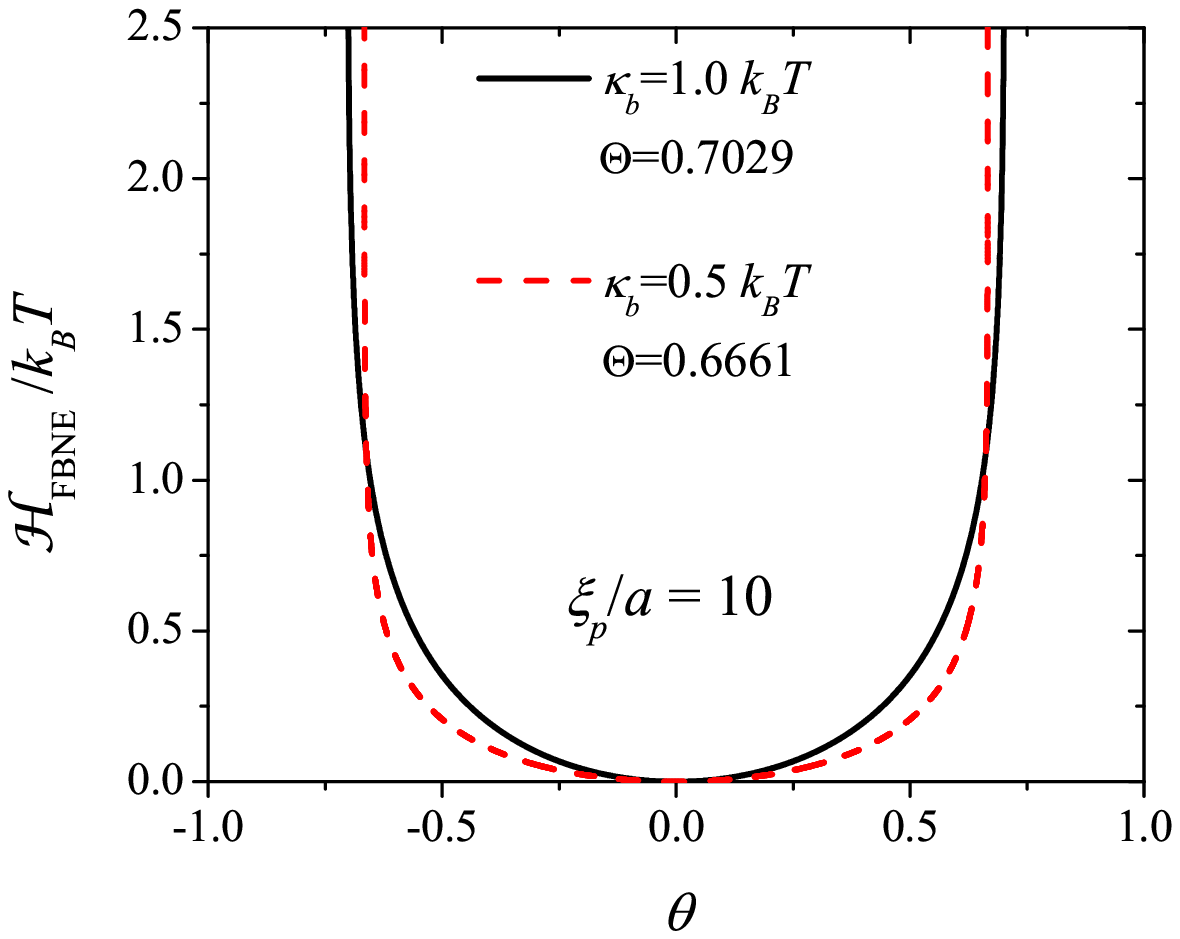}
    \label{fig:sub_FBNE}
  }
  \subfigure[]
  {
    \includegraphics[width=2.6in]{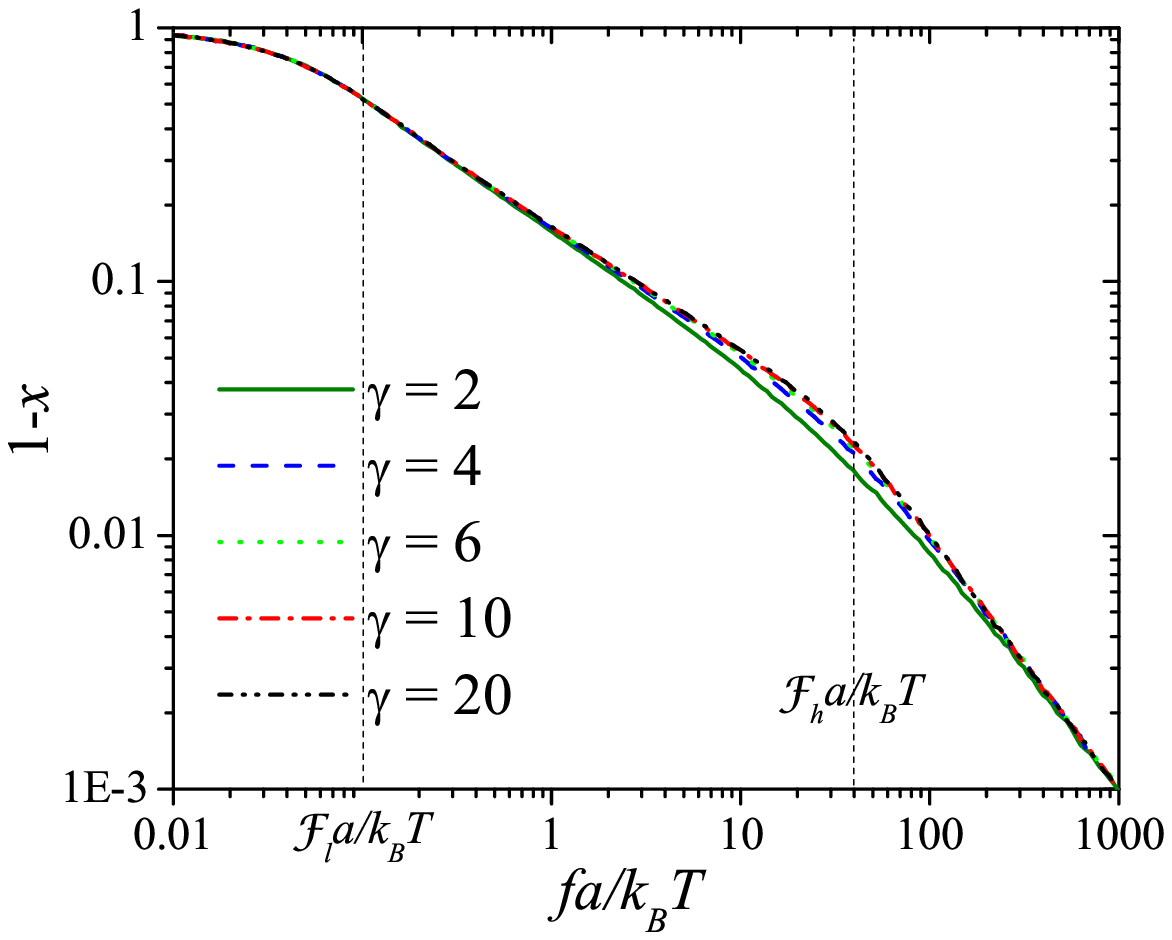}
    \label{fig:sub_log1xlog1fsimulations1}
  }
  \subfigure[]
  {
    \includegraphics[width=2.6in]{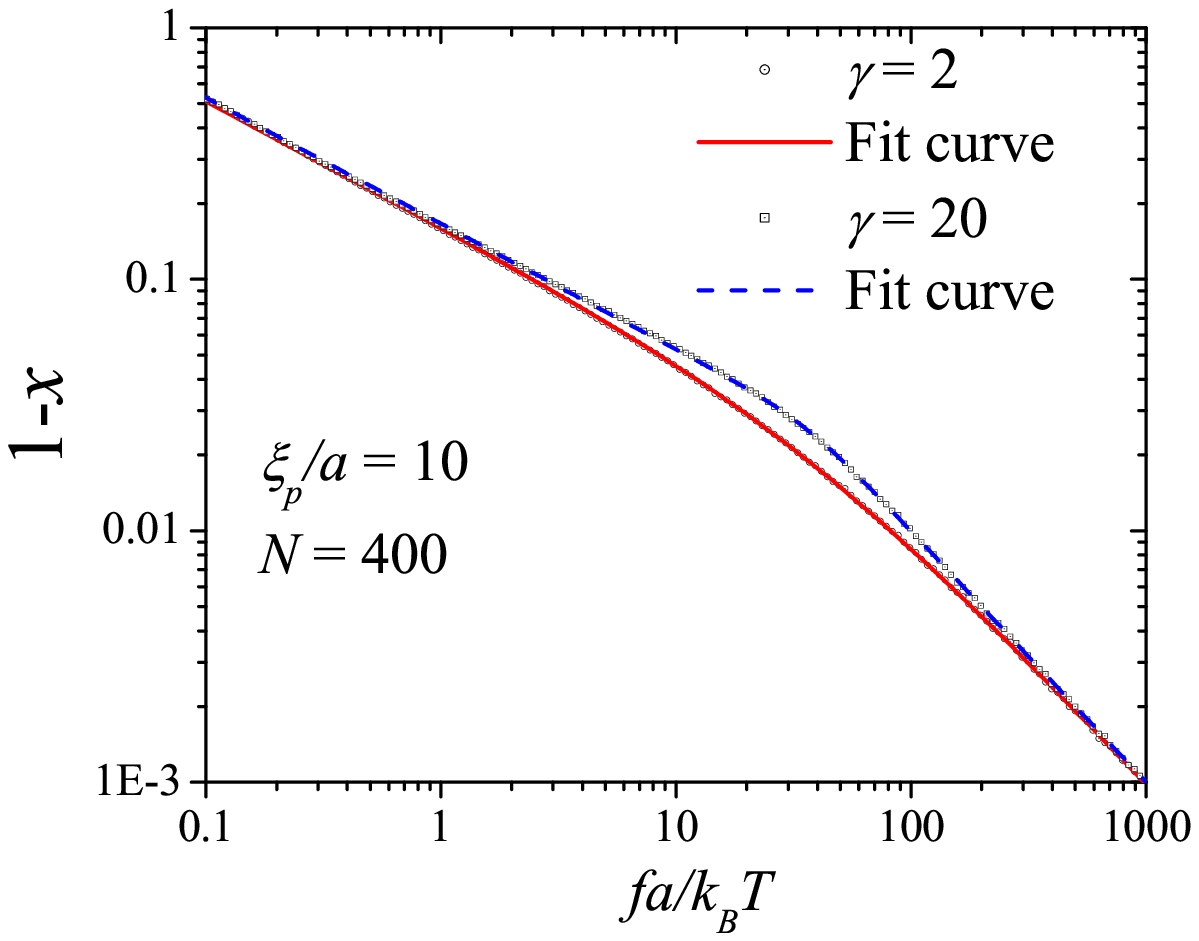}
    \label{fig:sub_log1xlog1ffit}
  }
  \caption{}
  \label{fig:simulations}
\end{figure}

{\bf Figure \ref{fig:simulations}}. Simulation of model Hamiltonians and fits to the simulation data. (a) Plot of ${\cal H}_b/\kappa_b$ as a function of $\theta/\Theta$ at various values of $\gamma$ (Eq.~\eqref{eq:GeneralizedWLC_Hamiltonian}). As $\gamma$ increases from 2, which is approximately the value for the WLC model, the potential mimics the excluded volume interaction between consecutive monomers. (b) Plot of ${\cal H}_{FBNE}(\theta)/k_BT$ in Eq.\eqref{eq:FBNE_Hamiltonian} at two sets of values of $\kappa_b$ and $\Theta$ with $\xi_p/a = 10$. (c) Log-log plot of $1-x$ as a function of $fa/k_BT$ obtained from Monte Carlo simulations for the polynomial model with $\xi_p/a = 10$ and $N = 400$ and various values of $\gamma$. The scaling laws $1 - x\sim f^{-1/2}$ and $1 - x\sim f^{-1}$, as well as the two crossover forces $\fl$ and $\fh$ are clearly visible. (d) The fits of Eq.~\eqref{eq:general_1S_2S} to the simulation data for $\gamma = 2$ and 20.

\begin{figure}[h]
  \centering
  \subfigure[]
  {
    \includegraphics[width=3in]{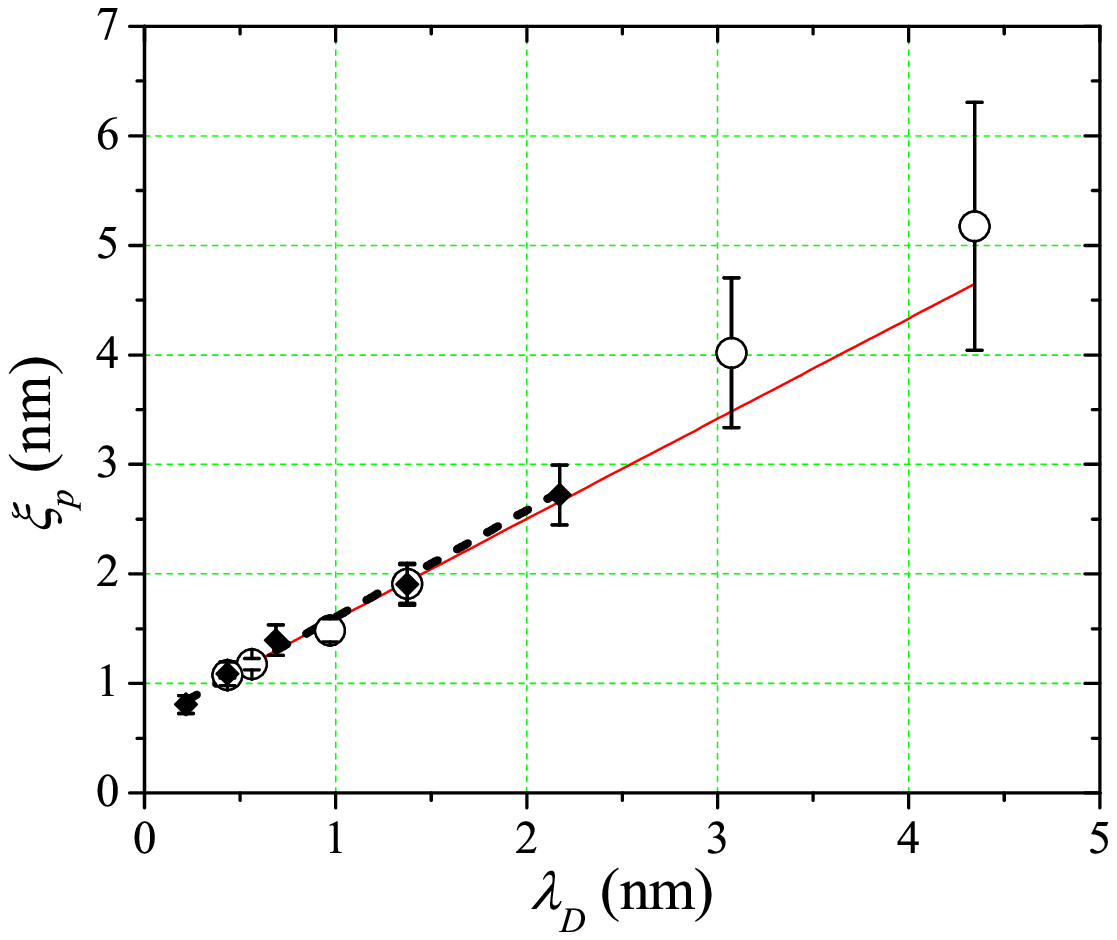}
    \label{fig:fit_ssRNA_DNA}
  }
  \subfigure[]
  {
    \includegraphics[width=3in]{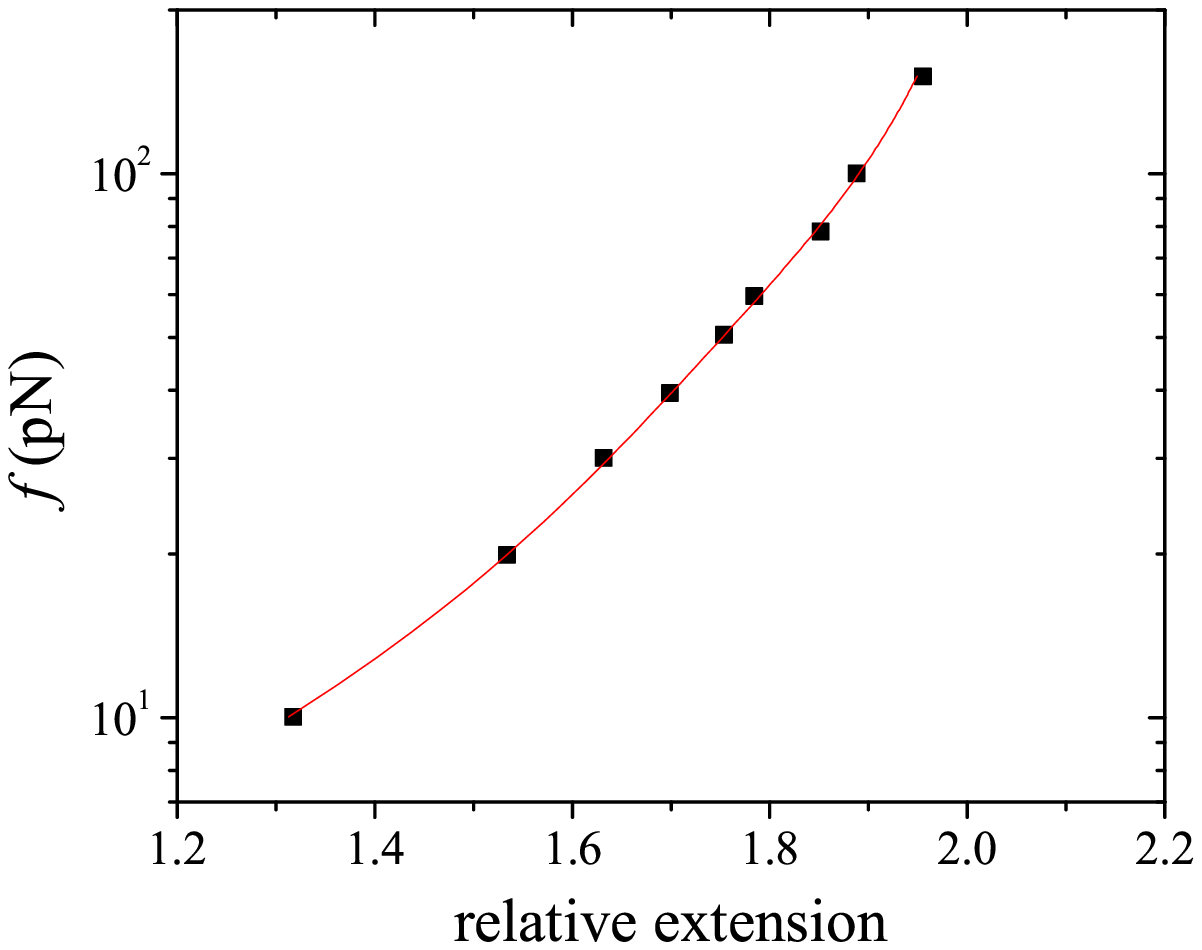}
    \label{fig:fit_dessinges}
  }
  \caption{}
  \label{fig:fit_koen_saleh_dessinges}
\end{figure}

{\bf Figure \ref{fig:fit_koen_saleh_dessinges}}. Analysis of the experimental results. (a) Estimates of $\xi_p(\lambda_D)$ for poly-U (circles) and ssDNA (filled symbols) using the $1/2$-rule. Linear fits to the data (solid and dashed lines) yielded the bare persistence lengths $\xi_p^{0,\text{poly-U}} \approx 0.67$ nm and $\xi_p^{0,\text{ssDNA}} \approx 0.63$. (b) The fit of equation~\eqref{eq:general_1S_2S} (solid line) to the FEC of a ssDNA (squares) by Rief et al.~\cite{MRief}.The extracted values for the parameters are: $\xi_p \approx 0.8$nm, $a \approx 0.5$ nm and $\alpha \approx 7$ leading to $\fl \approx 5$ pN and $\fh \approx 55$ pN. The extracted values of $\xi_p$ and $a$ are consistent with those reported in ref.~\cite{Rivetti98} and~\cite{Hyeon2006}, respectively. The value of $\alpha \approx 7$ implies that $\gamma$ (see Eq.~\ref{eq:GeneralizedWLC_Hamiltonian}) is large. As a result, the bending energy increases more sharply than predicted by the WLC model (Fig.~\ref{fig:sub_PP}) as $\theta$ increases beyond a critical value.

\begin{figure}[h]
  \centering
  \includegraphics[width=4in]{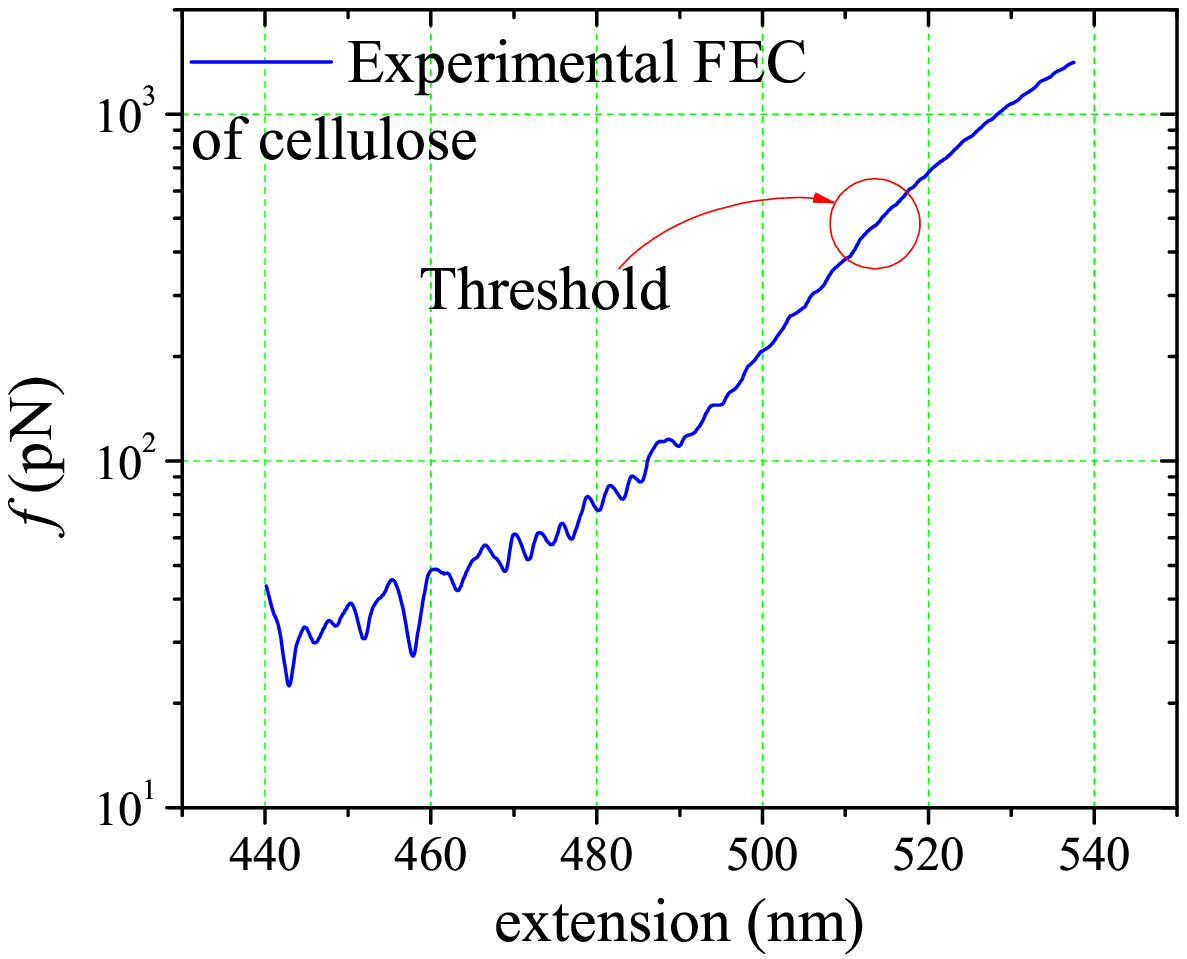}
  \caption{}
  \label{fig:threshold_detection}
\end{figure}

{\bf Figure \ref{fig:threshold_detection}}. An illustration of overstretching in cellulose. A threshold is identified as a narrow region in the circle, where it shows a decrease in the slope of the FEC when plotted in log-linear scale.

\newpage

\newpage

\newpage

\newpage

\end{document}